\documentclass{sig-alternate-2013}
\usepackage{graphicx}
\usepackage{amsmath}
\usepackage{multirow}

\usepackage{url}
\usepackage{float}
\usepackage{color}
\usepackage{enumerate}

\newcommand{\ignore}[1]{}

\newcommand{\ya}{\textsc{\emph{YA}}}

  \makeatletter
  \let\@copyrightspace\relax
  \makeatother
\begin{document}
%



\title{The Social World of Content Abusers\\ in Community Question Answering}
\subtitle{[Please cite the WWW'15 version of this paper]}

%
%
%
%
%

\numberofauthors{5} 
%
\author{
%
%
\alignauthor
Imrul Kayes\\
       \affaddr{University of South Florida}\\
       \affaddr{Tampa FL, USA}\\
       \email{imrul@mail.usf.edu}
\and
Nicolas Kourtellis\\
       \affaddr{Yahoo Labs}\\
       \affaddr{Barcelona, Spain}\\
       \email{kourtell@yahoo-inc.com}
\and
Daniele Quercia\\
       \affaddr{Yahoo Labs}\\
       \affaddr{Barcelona, Spain}\\
       \email{dquercia@yahoo-inc.com}
\and
Adriana Iamnitchi\\
   	\affaddr{University of South Florida}\\
    	\affaddr{Tampa FL, USA}\\
       	\email{anda@cse.usf.edu}
\and
Francesco Bonchi\\
       \affaddr{Yahoo Labs}\\
       \affaddr{Barcelona, Spain}\\
       \email{bonchi@yahoo-inc.com}
}


\maketitle

\begin{abstract}

Community-based question answering platforms can be rich sources of information on a variety of specialized topics, from finance to cooking. 
The usefulness of such platforms depends heavily on user contributions (questions and answers), but also on respecting the community rules. 
As a crowd-sourced service, such platforms rely on their users for monitoring and flagging content that violates community rules. 

Common wisdom is to eliminate the users who receive many flags. 
Our analysis of a year of traces from a mature Q\&A site shows that the number of flags does not tell the full story: on one hand, users with many flags may still contribute positively to the community. 
On the other hand, users who never get flagged are found to violate community rules and get their accounts suspended. 
This analysis, however, also shows that abusive users are betrayed by their network properties: we find strong evidence of homophilous behavior and use this finding to detect abusive users who go under the community radar. 
Based on our empirical observations, we build a classifier that is able to  detect abusive users with an accuracy as high as 83\%.



\end{abstract}

\section{Introduction}\label{sec:intro}

Community-based Question-Answering (CQA) sites, such as Yahoo Answers, Quora and Stack Overflow, are now rich and mature repositories of user-contributed questions and answers.
For example, Yahoo Answers (\ya), launched in December 2005, has more than one billion posted answers,\footnote{http://www.yanswersbloguk.com/b4/2010/05/04/1-billion-answers-served/} and Quora, one of the fastest growing CQA sites has seen three times growth in 2013.\footnote{http://www.goo.gl/MfK83y}

Like many other Internet communities, CQA platforms define community rules and expect users to obey them.
To enforce these rules, published as community guidelines and terms of services, these platforms provide users with tools to flag inappropriate content. 
In addition to community monitoring, some platforms employ human monitors to evaluate abuses and determine the appropriate responses, from removing content to suspending user accounts. 

To the best of our knowledge, this paper is the first to investigate the reporting of rule violations in \ya, one of the oldest, largest, and most popular CQA platforms.
The outcomes of this study could aid human monitors with automated tools in order to maintain the health of the community.
Our sampled dataset contains $10$ million editorially curated abuse reports posted between 2012 and 2013, and $1.5$ million users who submitted content during the one-year observation period, with about $9$\% of the users having their accounts suspended.
We use suspended accounts as a ground truth of bad behavior in \ya, and we refer to these users as \emph{content abusers}. 

We discover that, although used correctly, flags do not tell accurately which users should be suspended: while 32\% of the users active in our observation period have at least one flag, only 16\% of them are suspended during this time. 
Even considering the top 1\% users with the largest number of flags, only about 50\% of them deserve account suspension. 
Moreover, we see that users with lots of flags contribute positively to the community in terms of providing (even best) answers. 
Complicating an already complex problem, we find that $40$\% of the suspended users have not received any flags.

To reduce this large gray area of questionable behavior, we employ social network analysis tools in an attempt to understand the position of content abusers in the \textit{YA}  community. 
We learned that the follower-followee social network tunnels user attention not only in terms of generating answers to posted questions, but also in monitoring user behavior. 
More importantly, it turns out that this social network divulges information about the users who go under the community radar and never get flagged even if they seriously violate community rules.
This network-based information, combined with user activity, leads to accurate detection of the ``bad guys'': our classifier is able to distinguish between suspended and fair users with an accuracy as high as $83$\%.

The paper is structured as follows. 
Section~\ref{sec:related} discusses previous analysis of CQA platforms and the existing body of work on unethical behavior in online communities in general.
Section~\ref{sec:ya-details-datasets} presents the \emph{YA} functionalities relevant to this study and the dataset used. 
We introduce  a deviance score in Section~\ref{sec:flagging} that identifies the pool of bad users more accurately than the number of flags alone.  
Section~\ref{sec:deviance-contribution} demonstrates that deviant users are not all bad: despite their high deviance score, in aggregate their presence in the community is beneficial.
Section~\ref{sec:behavior-coordination} shows the effects of the social network on user contribution and behavior. 
Section~\ref{sec:classification} shows the classification of suspended and fair users.
We discuss the impact of these results in Section~\ref{sec:discussion}.

\section{Related Work}\label{sec:related}

We collate  past research on Community-based Question Answering (CQA) in five categories depending on whether it has dealt with content, users, new applications, bad behavior in online settings, or CQA communication networks.

\textbf{Content.} Research in this area has investigated textual aspects  of questions and answers. In so doing, it has proposed algorithmic solutions to automatically determine: the quality of questions~\cite{Li2012Question,Sun2009Questions} and answers~\cite{Shah2010Answers,Agichtein2008Answers}, the extent to which certain questions are easy to answer~\cite{dror2013will,Richardson2011Answerability}, and the type of a given question (e.g., factual or conversational)~\cite{harper2009facts}.
\noindent

\textbf{Users.} Research on CQA users has  been mostly about understanding why users contribute content: that is, why users ask questions (askers are failed searchers, in that, they use CQA sites when web search fails~\cite{Liu2012WSF}); and why they answer questions (e.g., they refrain from answering sensitive questions to avoid being  reported for abuse and potentially lose access to the community~\cite{Dearman2010Why}).
\noindent

\textbf{New applications.} As for applications, research has proposed effective ways of recommending questions to the most appropriate answerers 
~\cite{Qu2009PQR,Szpektor2013Reco}, of automatically answering questions based on past answers~\cite{Shtok2012LPA}, and of retrieving factual answers~\cite{Bian2008Finding} or factual bits within an answer~\cite{Weber2012ALE}.
\noindent

\textbf{Bad behavior in online settings.} Qualitative and quantitative studies of bad behavior in online settings have been done before including newsgroups~\cite{phillips1996defending}, online chat communities~\cite{suler1998bad}, and online multiplayer video games~\cite{Blackburn2012BSC}.
A body of work also investigates the impact of the bad behavior.
Researchers find that bad behavior has negative effects on the community and its members: it decreases community's cohesion~\cite{wellen2006deviance}, performance~\cite{dunlop2004workplace} and participation~\cite{davis2002deviance}.
In the worst case, users who are the targets of bad behavior may leave or avoid online social spaces~\cite{davis2002deviance}.
\noindent

\textbf{Communication networks.} The communication networks behind CQA sites have been recently studied. More specifically, researchers have explored the relationship between  content quality and network properties such as number of followers~\cite{Wang2013WSC} and  tie strength~\cite{Panovich2012CQA}.
\noindent

Research on CQA communication networks is quite recent, so it comes as no surprise that there has not been any work on how such networks mediate different types of behavior on CQA sites.
This paper, for the first time, sheds light on bad behavior in CQA communities by studying \ya, one of the largest and oldest such communities.
It quantifies how \ya's networks channel user attention, and how that results in different behavioral patterns that can be used to limit bad behavior.\\

\section{Yahoo Answers}\label{sec:ya-details-datasets}

After $9$ years of activity, \emph{YA}  has $56$M monthly visitors (U.S. only).\footnote{http://www.listofsearchengines.org/qa-search-engines}
The functionalities of the \emph{YA} platform and the dataset used in this analysis are presented next.

\subsection{The Platform}

\emph{YA} is a CQA platform in which community members ask and answer questions on various topics.
Users ask questions and assign them to categories selected from a predefined taxonomy, e.g., \emph{Business \& Finance}, \emph{Health}, and  \emph{Politics \& Government}.
Users can find questions by searching or browsing through this hierarchy of categories.
A question has a title (typically, a short summary of the question), and a body with additional details. 

A user can answer any question but can post only one answer per question.
Questions remain open for four days for others to answer.
However, the asker can select a best answer before the end of this 4-day period, which automatically \emph{resolves} the question and archives it as a \emph{reference} question.
The best answer can also be rated between one to five, known as \emph{answer rating}.
If the asker does not choose a best answer, the community selects  one through voting.
The asker can extend the answering duration for an extra four days.
The questions left unanswered after the allowed duration are deleted from the site.
In addition to questions and answers, users can contribute comments to questions already answered and archived.

\emph{YA} has a system of points and levels to encourage and reward participation.\footnote{https://answers.yahoo.com/info/scoring\_system}
A user is penalized five points for posting a question, but if she chooses a best answer for her question, three points are given back.
A user who posts an answer receives two points; a best answer is worth $10$ points.
A leaderboard, updated daily, ranks users based on the total number of points they collected.
Users are split into seven levels based on their acquired points (e.g., 1-249 points: level 1, 250-999 points: level 2, ..., 25000+ points: level 7).
These levels are used to limit user actions, such as posting questions, answers, comments, follows, and votes: e.g., first level users can ask $5$ questions and provide $20$ answers in a day.

\emph{YA} requires its users to follow the Community Guidelines that forbids users to post spam, insults, or rants, and the Yahoo Terms of Service~\cite{YahooAnswersCommunity} that limits harm to minors, harassment, privacy invasion, impersonation and misrepresentation, and fraud and phishing. 
Users can flag content (questions, answers or comments) that violates the Community Guidelines and Terms of Service using the ``Report Abuse'' functionality.
Users click on a flag sign embedded with the content and choose a reason between violation of the community guidelines and violation of the terms of service.
Reported content is then verified by human inspectors before it is deleted from the platform.

Users in \emph{YA} can choose to follow other users, thus creating a follower-followee relationship used for information dissemination. 
The followee's actions (e.g., questions, answers, ratings, votes, best answer, awards) are automatically posted on the follower's newsfeed. 
In addition, users can follow questions, in which case all responses are sent to the followers of that question.\\\\

\subsection{Dataset}

\begin{figure}[htbp]
\centering
\includegraphics[height=3cm]{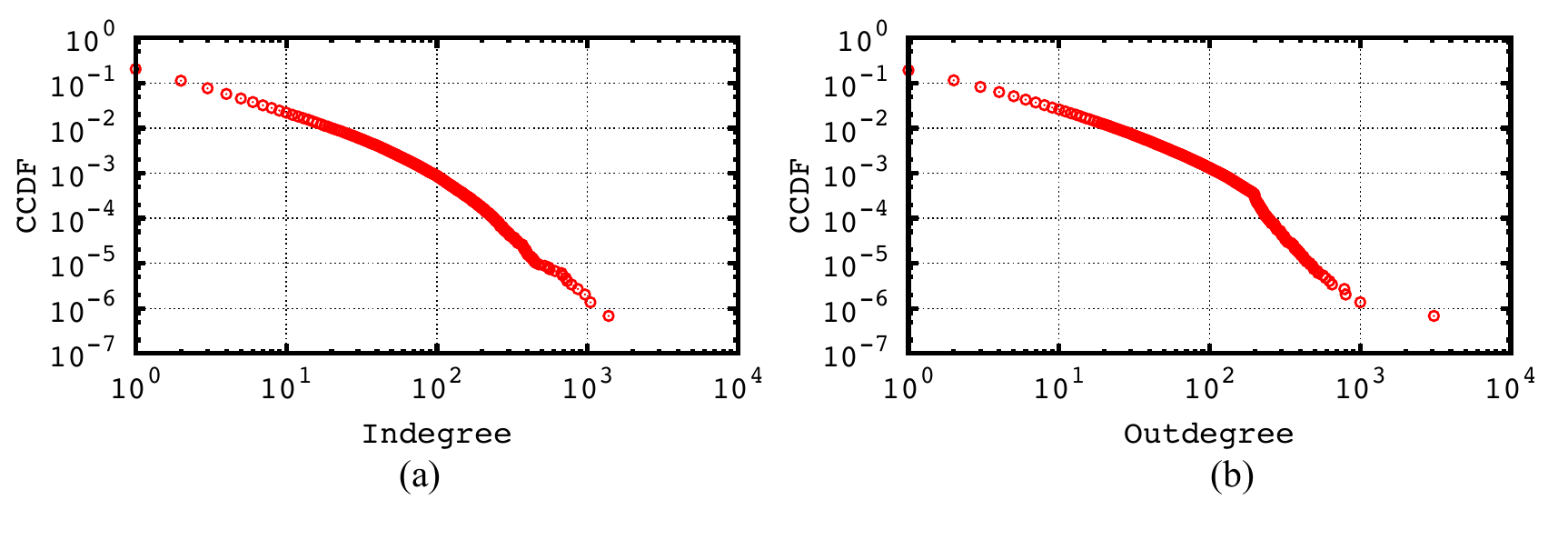}
\caption{(a) Indegree distribution; (b) Outdegree distribution.}
\label{fig:in_out_degree}
\end{figure}

We studied a sample of 10 million abuse reports posted between 2012 and 2013 originating from 1.5 million active users.
These users are connected via $2.6$ million follower-followee relationships in a social network (referred to as $FF$ in this study) that has $165,441$ weakly connected components.
The largest weakly connected component has  $1.1$M nodes ($74.32$\% of the nodes) and $2.4$M edges ($91.37$\% of the edges).
Out of the 1.5 million users, about $9$\% of the users have been suspended from the community.
Figure~\ref{fig:in_out_degree}(a) and Figure~\ref{fig:in_out_degree}(b) plot the complementary cumulative distribution function (CCDF) for the degree of followers (indegree) and followees (outdegree), respectively.
The indegree and outdegree follow power-law distributions~\cite{barabasi1999emergence}, with an exponential fitting parameter $\alpha$ $3.53$ and $2.95$ respectively.

Along with the follower-followee social network, we built an activity network ($AN$) that connects users if they interacted with each other's content.
In the \emph{AN} network, nodes are users who answered other users' questions, directed edges point from the answerer to the asker.
The activity network has  $1.2$M nodes and $45$M edges, thus being $141$ times denser (ratio of the number of edges to the number of possible edges) than the $FF$ network.

\section{Flagging in Yahoo Answers}\label{sec:flagging}

In this section, we study whether flags (we use flags and abuse reports interchangeably) can be used as an appropriate proxy for content abuse.
First, we investigate whether the flags reported from users are typically valid, i.e. if human inspectors remove the flagged content and further, how quickly this is done (Section~\ref{abuse-reports}).
Then, we explore how the flags can be used to detect content abusers (Sections~\ref{deviant-users} and~\ref{deviance-score-suspension}).

\subsection{Abuse Reports} \label{abuse-reports}

\emph{YA} is a self-moderating community; the health of the platform depends on community contributions in terms of reporting abuses.
Besides participating by providing questions and answers, \emph{YA} users also contribute to the platform by reporting abusive content.
Reporters serve as an intermediate layer in the \emph{YA} moderation process since these abuse reports are verified by human inspectors.
If the report is valid, the content is promptly deleted. 

\begin{figure}[htbp]
\centering
\includegraphics[scale=0.7]{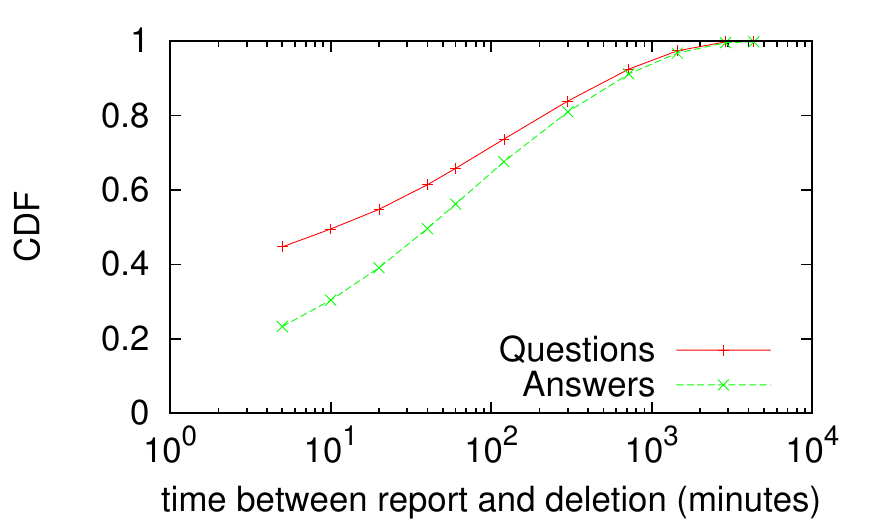}
\caption{The CDF of the time delay between the posting of the content (questions or answers) and its deletion due to valid abuse reporting.}
\label{fig:content_deleted}
\end{figure}

To check if valid abuse reports are indeed an accurate sensor for the correct monitoring of the platform, we look at how soon a report is curated. 
Figure~\ref{fig:content_deleted} shows the distributions of the time interval between the time when a content (question or answer) is posted and when it is deleted due to abuse reports. 
About $97$\% of questions and answers marked as abusive are deleted within the same day they are posted.
All reported abusive questions and answers are deleted within three days of posting.

This result highlights two facts.
First, the users monitoring the platform act very quickly on content: within 10 minutes from being posted, 50\% of the bad posts are reported. 
Second, the validation of abuse reports happens within 3 days (and in vast majority within a day). 
Hence, in our dataset, if there are abuse reports that did not have the chance of being curated yet and thus we do not consider them, those are too few to impact our analysis. 

However, the abuse reporting functionality might be abused as well, due to several reasons.
First, reporting is an easy and fast process, requiring only a few steps. 
Second, a user is not penalized for misreporting content abuse, perhaps in an attempt to not discourage users from exercising good citizenship.
And third, independent of their level in the \emph{YA} platform (that limits the number of questions and answers posted per day), users can report an unlimited number of abuses. 

\begin{figure}[htbp]
\centering
\includegraphics[height=5cm]{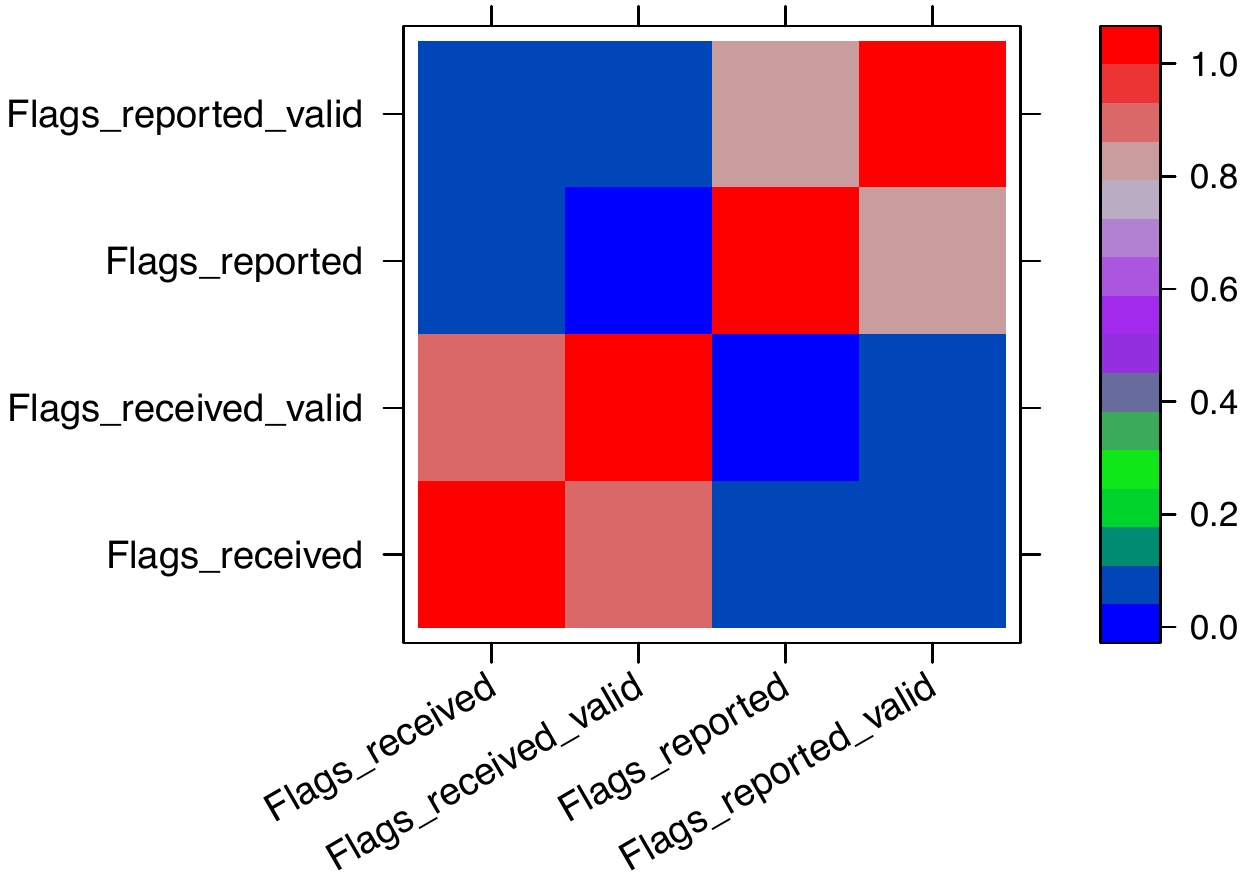}
\caption{The Pearson correlation coefficient heat map of flags received, valid flags received, flags reported and valid flags reported on questions. All values are statistically significant ($p$-values \textless $0.01$).}
\label{fig:flags_heatmap_question}
\end{figure}

To check whether users abuse the abuse reporting functionality, we compare the number of flags received/reported with the number of validated flags received/reported per user.
Figure~\ref{fig:flags_heatmap_question} shows a correlation heat map of the flags received and flags received that are valid, as well as flags reported and flags reported that are valid, on questions and for all contributors (results on answers are similar and are excluded for brevity).
For questions (answers), we have a very high correlation between flags received by users and flags that are valid ($r=0.90$ ($0.87$), $p<0.01$) and between flags reported by users and that are valid ($r=0.80$ ($0.92$), $p<0.01$).

These high correlations indicate that, in general, users are not exploiting the abuse reporting functionality.
When a user reports an abuse, it is very likely that the content is violating community rules.
Another interesting finding from the correlation heat maps is that for both questions and answers, users have almost negligible or very weak correlation between the number of  flags they reported that are valid and the number of flags they received that are valid. 
This hints that the good guys of the community are not bad guys at the same time: the users who correctly report a lot of content abuses are not posting abusive content themselves. 

\subsection{Deviant Users}\label{deviant-users}

Given that flags are good proxies for identifying bad content, how should they be used to detect content abusers and thus determine which accounts to be suspended?
Common wisdom might suggest that content abusers are those who receive a large number of flags. 
Of the top $1$\% flagged askers and answerers, we find $51.63$\% and $53.89$\%, respectively, are suspended.
But finding a threshold on the number of flags received by a user is not likely to work accurately for content abuser detection: users with low activity who received flags for all their posts might go below this threshold. 
At the same time, highly active users may collect many flags even if for a small percentage of their posts, yet contribute significantly to the community.

\begin{figure}[htbp]
\centering
\includegraphics[height=3.3cm]{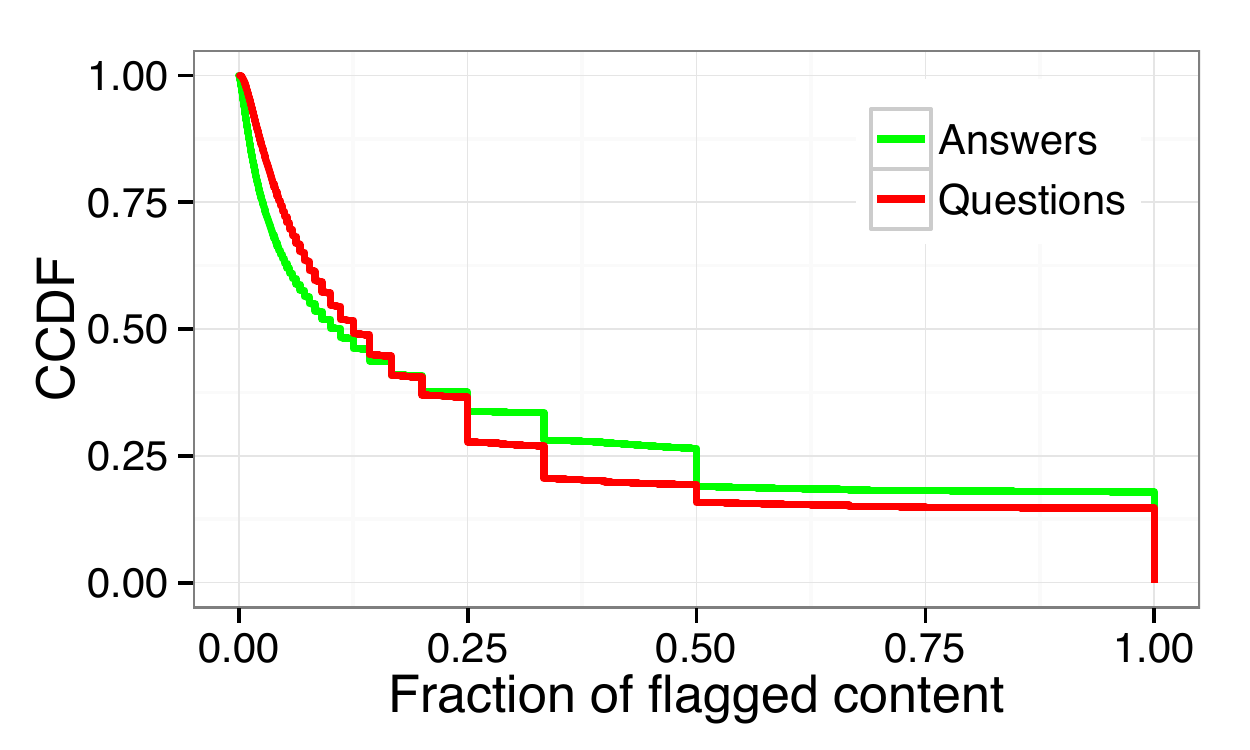}
\caption{Distributions of fraction of flagged questions and answers.}
\label{fig:fraction_flags_distribution}
\end{figure}

This intuition motivated us to measure the correlation between a user's number of posts and the number of flags received. 
Indeed, we find that the correlation between the number of questions a user asks and the number of valid flags she receives from others is  high ($r=0.49$, $p<0.05$).
Similarly, the number of answers posted and the number of valid flags received per user are highly correlated ($r = 0.37$, $p<0.05$).
The distributions of the fraction of flagged questions and answers is shown in Figure~\ref{fig:fraction_flags_distribution}.
While about $27$\% users have more than $25$\% flagged questions, about $34$\% users have more than $25$\%  flagged answers.
Also, about $16$\% and $19$\% of users have more than $50$\%  flagged questions and answers respectively.

So, instead of directly considering flags, we define a \emph{deviance score} metric that indicates how much a user deviates from the norm in terms of received flags considering the amount of her activity.
Deviant behavior is defined by actions or behaviors that are contrary to the dominant norms of the society~\cite{douglas1982sociology}.
Although social norms differ from culture to culture, within a context, they remain the same and they are the rules by which the members of the community are conventionally guided.

We define the deviance score for a user $u$ as the number of correct abuse reports (flags) she receives over the total content (question/answer) she posted, after eliminating the expected average number of correct abuse reports given the amount of content posted:
\begin{equation}
\centering
\begin{split}
 \textrm{Deviance}_{\textrm{Q/A}}(u)=  Y_{Q/A,u}-\hat{Y}_{Q/A,u} \end{split}
\label{eq:deviance_QA}
\end{equation} 
where $Y_{Q/A,u}$ is the number of correct abuse reports received by $u$ for her questions/answers, and $\hat{Y}_{Q/A,u}$ is the expected number of correct abuse reports to be received by $u$ for those questions/answers.

To capture the expected number of the correct abuse reports a user receives for questions/answers, we considered a number of linear and polynomial regression models between the response variable (number of correct abuse reports) and the predictor variable (number of questions/answers) for all users.
Among them, the following linear model was the best in explaining the variability of the response variable.
\begin{equation}
\centering
\begin{split}
 Y= \alpha+ \beta X + \epsilon
 \end{split}
\end{equation} 
where $Y$ is the number of correct abuse reports (flags) received for the content, $X$ is the number of content posts and $\epsilon$ is the error term.

In eq.~(\ref{eq:deviance_QA}), a positive deviance score reflects deviant users, i.e., those whose deviance cannot be only explained by their activity levels.

\subsection{Deviance Score vs. Suspension}\label{deviance-score-suspension}
\begin{figure}[htbp]
\centering
\includegraphics[height=2.75cm]{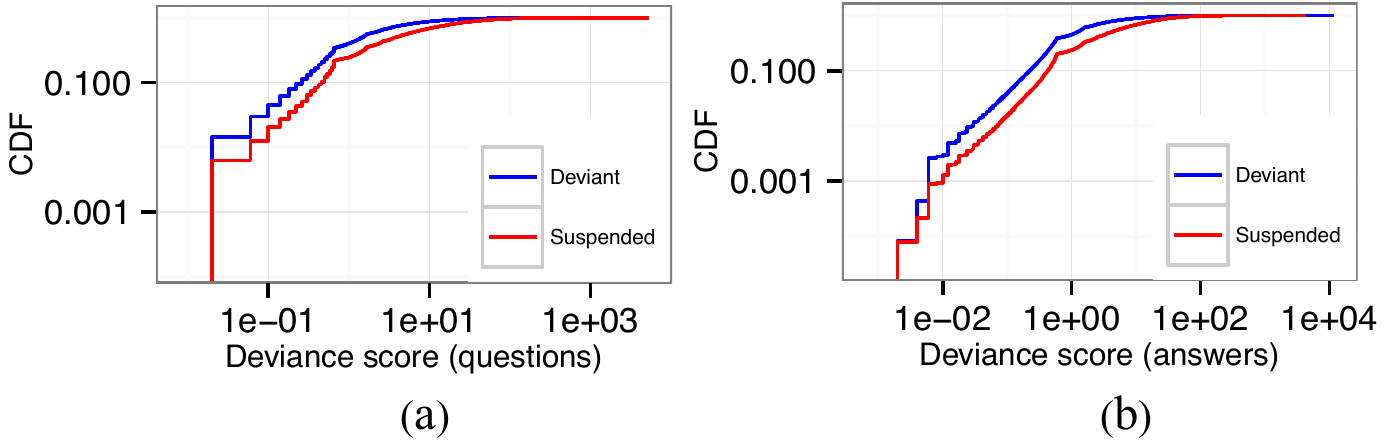}
\caption{The CDF of suspended and deviant users' deviance scores for (a) questions; (b) answers.
Distributions are different with $p$<$0.001$ for both KS and permutation tests
(for questions: $D = 0.22$, $Z =46.04$ and
for answers: $D = 0.28$, $Z =50.53$.)}
\label{fig:cdf-suspended-deviant-deviance}
\end{figure}

We found $105,340$ users with positive \emph{question} deviance scores and $121,705$ users with positive \emph{answer} deviance scores.
Among the users with positive question deviance score, $31,891$ users ($30.27$\%) have been suspended.
Similarly, among the users with a positive answer deviance score, $37,633$ users ($30.92$\%) have been suspended.
The CDF  of suspended and deviant (but not suspended) users' deviance scores for both questions and answers is shown in Figure~\ref{fig:cdf-suspended-deviant-deviance}.
In both cases, suspended and deviant users are visibly characterized by different distributions: suspended users tend to have higher deviance scores than deviant (not suspended) users.
While this difference is visually apparent, we also ensure it is statistically significant using two methods: 1) the two-sample Kolmogorov-Smirnov (KS) test, and 2) a permutation test, to verify that the two samples are drawn from different probability distributions.

\begin{figure}[htbp]
\centering
\includegraphics[scale=0.35]{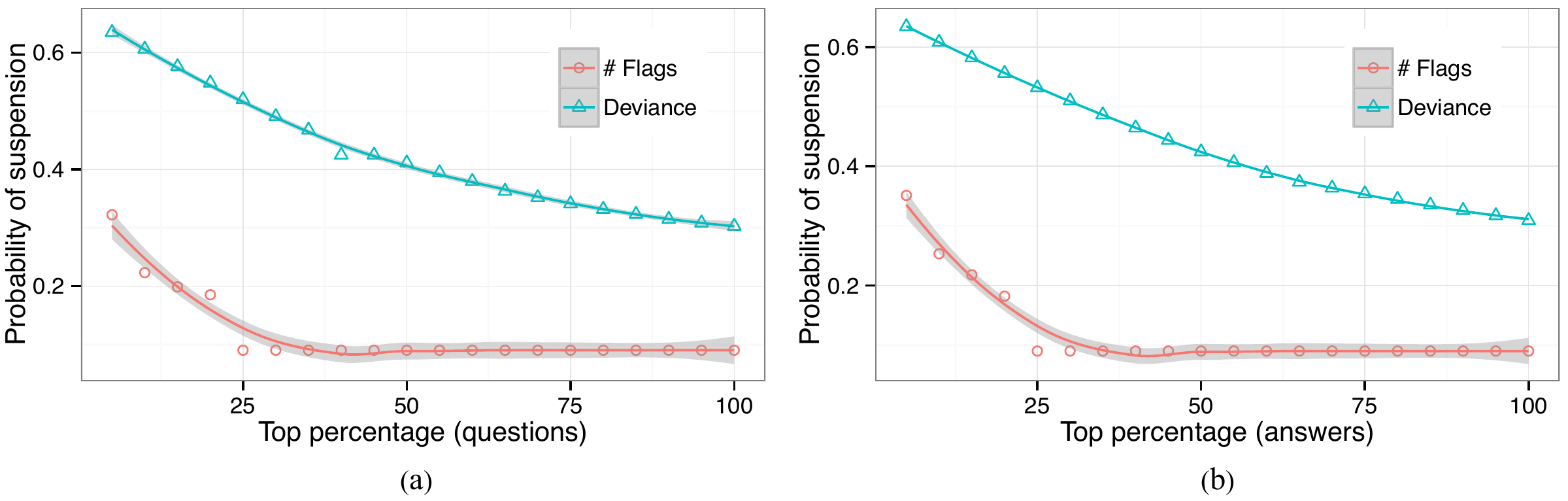}
\caption{Probability of being suspended, given a user is within top $x$\% of (a) question or (b) answer deviance scores and flags. Local polynomial regression fitting with 95\% confidence interval area is also shown.}
\label{fig:prob_suspension}
\end{figure}

We also find that $63.94$\% of top $1$\% deviant question askers' and $64.77$\% of top $1$\% deviant answerers' accounts have been suspended.
This hints that the higher deviance score a user has, the more likely (s)he is to be removed from the community. 
Figure~\ref{fig:prob_suspension} shows the probability of a user being suspended as a function of its rank in the community as expressed by deviance score and number of flags. 
We observe that the more deviant a user is, the more probable is that she will be suspended.
Also, in all cases, deviance score shows a higher probability of suspension compared to the number of flags. 

These results show that the deviance score is a better metric for identifying the content abusers than the number of flags is by itself.
However, both metrics fail to identify content abusers who go under the community radar. 
We found that about $40\%$ of the suspended users had never been flagged for the abusive content they certainly posted, thus maintaining a negative deviance score. 
Thus, our investigation into user behavior in the \emph{YA} community continues. 

\section{Deviant vs. Suspended Users} \label{sec:deviance-contribution}

Despite the fact that deviance score better identifies the pool of suspended users, it is clearly an imperfect metric.
On one hand, there are high deviance score users who are not suspended, despite the fact that the platform seems to be fairly quick in responding to abuse reports. 
On the other hand, there are ``ordinary'' users, according to the deviance score (i.e., with a negative deviance score) who are never reported for abusive content, yet get suspended. 
These users may even be fair users for a long time, but sometimes their posted content can be highly abusive (e.g., vulgar language and images) that platform moderators immediately suspend them.
To better understand these two groups of users---deviant but not suspended and suspended but not flagged---we analyze in more detail their activity. 
Note that the two groups are disjoint (i.e., deviant users have received at least one flag). 

\subsection{Deviance is Engaging}

One of the  success metrics of CQA platforms is \emph{user engagement}~\cite{Mamykina2011DLF}, which can be measured by the number of contributions and by the number of users who respond to a particular content. 
Thus, we use the number of answers deviant users receive to their questions and the number of distinct users who respond to the deviant users' questions as measures of deviants' contribution to user engagement with the platform. 
To this end, for each category of users (typical, deviant but not suspended, and suspended) we randomly selected $500k$ questions they asked. 
For each question, we extracted all answers received and also the users who answered those questions. 
Table~\ref{desAnswers} presents the statistics of the number of answers received per category of users.

\begin{figure}[htbp]
\centering
\includegraphics[scale=0.35]{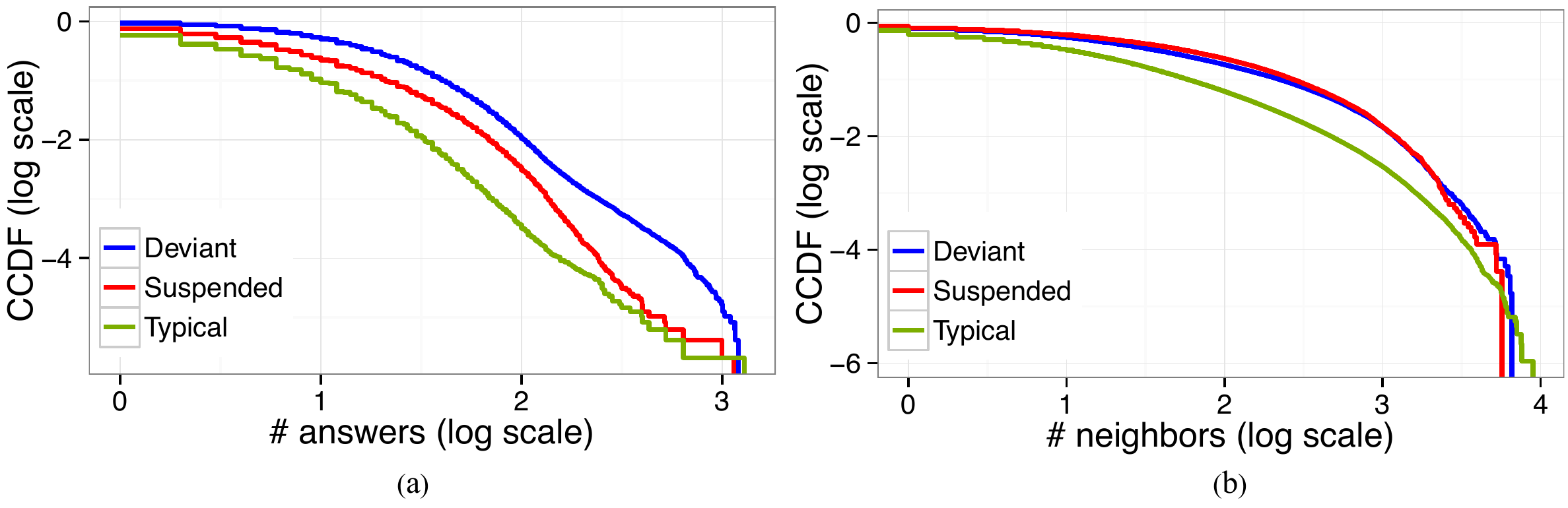}
\caption{ (a) CCDF of the number of answers received by the typical, deviant but not suspended, and suspended users on questions; (b) CCDF of the number of neighbors (distinct answerers) that typical, deviant but not suspended, and suspended users have.}
\label{fig:answerdiantfiar}
\end{figure}

Deviant users' questions get significantly more answers than typical users's questions get: on average, a question posted by a deviant user gets about 5 times more answers than the average question posted by a typical user. 
This difference is also seen in the CCDF of the number of answers received by typical, deviant and suspended users in Figure~\ref{fig:answerdiantfiar}(a).
The distributions (pairwise) are different with $p_{ks}<0.01$ and $p_{perm}<0.01$.

\begin{table}
\caption{Descriptive statistics of the number of answers received by typical, deviant but not suspended, and suspended users per question.}
\centering
\scalebox{0.85}{
\begin{tabular}{crrrrrrr}
    \hline
   Type&Min.& 1st Qu.&Med.&Mean&3rd Qu.&Max.\\
   \hline
Typical&  1.00 &1.00 &2.00&4.36&5.00&1296.00\\		
Deviant&1.00&5.00&11.00& 17.96& 22.00&1205.00\\
Suspended&1.00&1.00&4.00& 8.67& 9.00&1144.00\\
\hline
\end{tabular}
}
\label{desAnswers}
\end{table}

Deviant users not only attract more answers, but also interact with more users than typical users do, as shown by Figure~\ref{fig:answerdiantfiar}(b) and these two distributions are different ($p_{ks}<0.01$, $p_{perm}<0.01$).

This result from analyzing a random sample of $500k$ questions is confirmed when looking at the indegree of nodes in the activity network, which represents the number of users who answered that node's questions, as shown in Table~\ref{desAnswerers} for typical and deviant users.
Deviant askers have a higher number of neighbors than typical askers.
An explanation might be, as shown in~\cite{harper2009facts}, that users who ask conversational questions tend to have more neighbors (with whom the asker has interaction) than users who ask informational questions.
This suggests that deviant users tend to ask more conversational questions, which engage a larger number of responders. \\ \\

\begin{table}
\caption{Descriptive statistics of the number of neighbors askers have in the Activity Network.}
\centering
\scalebox{0.85}{
\begin{tabular}{crrrrrrr}
    \hline
   Type&Min.& 1st Qu.&Med.&Mean&3rd Qu.&Max.\\
   \hline
Typical&  0.00 &1.00 &5.00&28.16&19.00&13270.00\\		
Deviant&0.00&3.00&20.00& 103.40& 90.00&5698.00\\
Suspended&0.00&2.00&13.00& 88.62& 60.00&6576.00\\
\hline
\end{tabular}
}
\label{desAnswerers}
\end{table}

\subsection{Deviance is Noisy}

We observed that deviant users impact the \emph{quantity} of content in the system.
Do they impact \emph{quality}, too?
To address this question, we look at the percentage of the best answers with respect to the total number of answers submitted per user.



\begin{figure}[htbp]
\centering
	\includegraphics[scale=0.5]{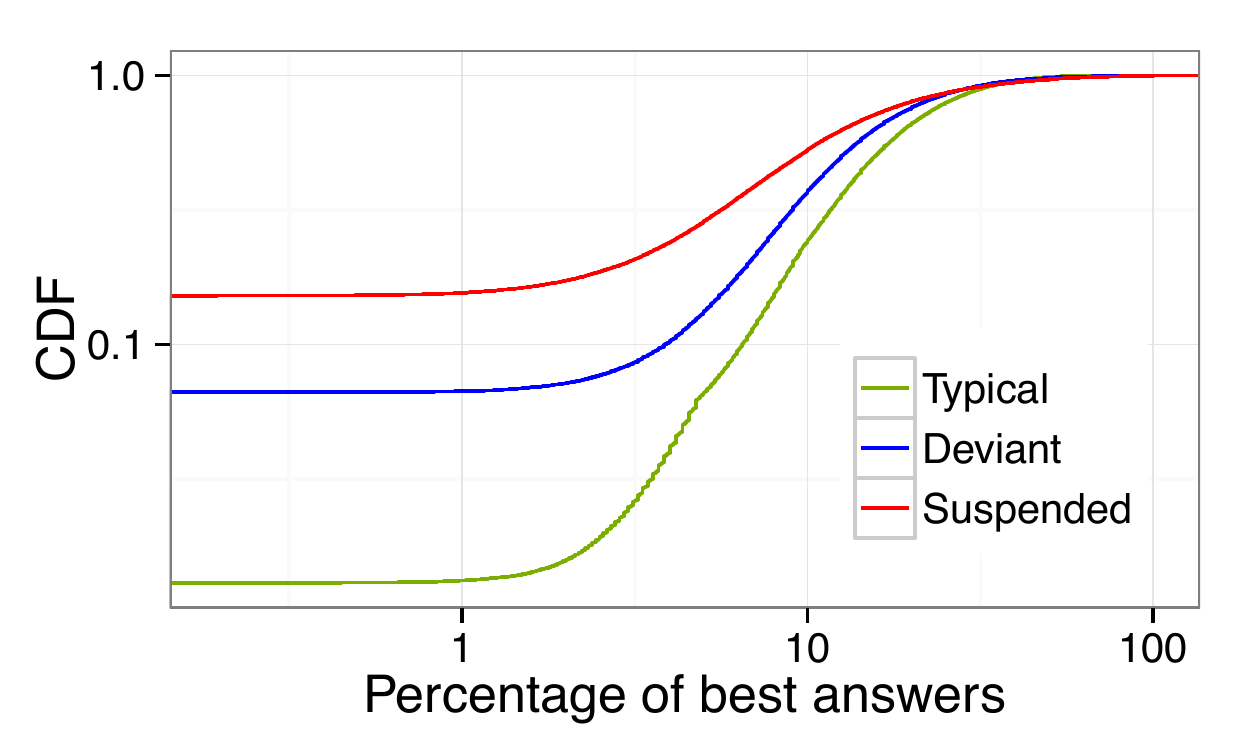}
\caption{ CDF of the percentage of best answers for  typical, deviant but not suspended and suspended users.}
\label{fig:best_answers_deviant_fair_suspended}
\end{figure}

Figure~\ref{fig:best_answers_deviant_fair_suspended} shows the CDF of the percentage of best answers for different classes of users: 1) typical, 2) deviant but not suspended, and 3) suspended. 
The results show that users who are moderately deviant but did not get suspended  have higher percentage of best answers than suspended users (distributions  are different $p_{ks}<0.01$, $p_{perm}<0.01$), but lower than that of typical users (distributions are different $p_{ks}<0.01$, $p_{perm}<0.01$).

To conclude, it turns out  that while deviant users are beneficial in terms of platform success metrics, as they increase user engagement by attracting more answers and attracting more users who answer their questions, they do not contribute more than the norm-following users in terms of content quality.

\subsection{The Suspended but Not Flagged Users}\label{sec:never-flagged-behavior}

While the results above show how the deviant users differ from the suspended and from the typical users, we do not have yet an understanding of the behavior of the users who get suspended without other users flagging their abusive content. 
An initial analysis of these users---suspended but not flagged---shows the following particularities when compared to the fair users (all users, independent of their deviance status, who are not suspended). 

\begin{figure*}[htbp]
\centering
\includegraphics[height=3.5cm]{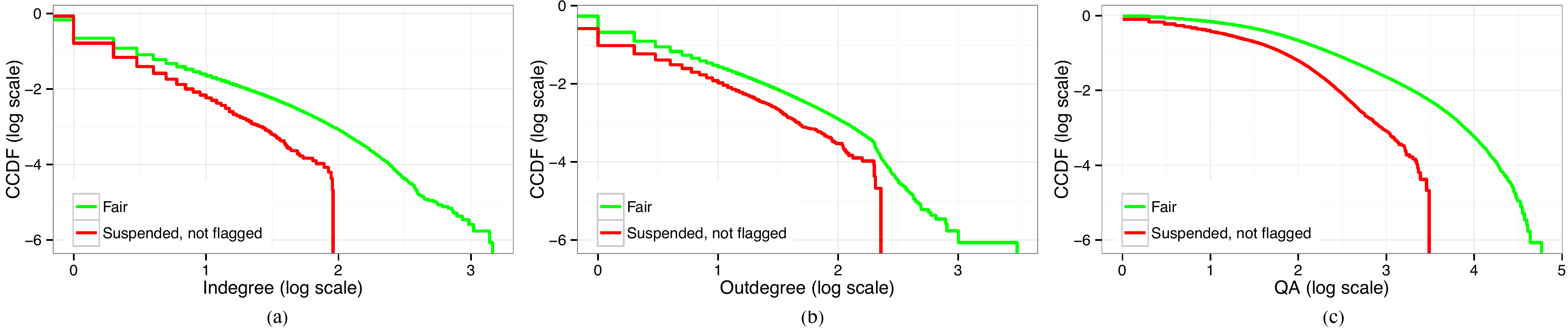}
\caption{Distributions of (a) indegree; (b) outdegree and (c) number of questions and answers (QA) of never flagged suspended users and fair users
(for outdegree: $D=0.28$ and $Z=27.40$, $p<0.001$,
for indegree: $D=0.17$ and $Z=15.86$, $p<0.001$
and for activity: $D=0.30$ and $Z=40.30$, $p<0.001$).
}
\label{fig:never-flagged-suspended}
\end{figure*}

First, they are followed by  and follow significantly fewer other users. 
Figures~\ref{fig:never-flagged-suspended}~(a) and (b) show the distributions of indegree and outdegree of never-flagged-suspended users compared to those of fair users.
Not only these users have smaller social circles, but they also have lower activity levels, as shown in Figure~\ref{fig:never-flagged-suspended}~(c).
Of course, these results could be correlated: low activity may mean low engagement in the social platform. 
These results may also suggest that (some of) these users join the platform for particular objectives that are orthogonal to the platform purpose, such as spamming. 
More importantly, however, these results suggest directions that we present in the following. \\\\\\

\section{Members of the Network}
\label{sec:behavior-coordination}

We investigate how the social network defined by the follower-followee relationships impacts user activities and behaviors in \ya.
Our final goal is to understand how to separate fair users from users who should be suspended even in the absence of flags. 
We learn that users close in the \emph{FF} network not only help each other by answering questions, but also monitor each other's behavior by reporting flags (Section~\ref{sec:answers-reports}).
Thus, the social network allows users to implicitly coordinate their behavior so much so that users who are socially close exhibit not only similar behavior, but also a similar deviation from the typical behavior (Section~\ref{sec:assortativity}).

\subsection{Out of Sight, Out of Mind}\label{sec:answers-reports}

We expect that users receive more answers from users that are close in the social network.
To verify this intuition, we randomly selected $7$M answers such that both parties of the interaction (the user who posted the question and the user who answered it) are in the social network, and measured the social distances between the two users.
For a user $u$ and a social distance $h$, the probability of receiving an answer from followers at distance $h$ is the following:

\begin{equation}
\resizebox{1.0\hsize}{!}{$
p_{h}=  \frac{\textrm{\# of $u$'s followers at distance $h$ who answered $u$'s questions}}{\textrm{\# of $u$'s followers at distance $h$}}
$}
\end{equation}

\begin{figure}[htbp]
\centering
\includegraphics[height=4cm]{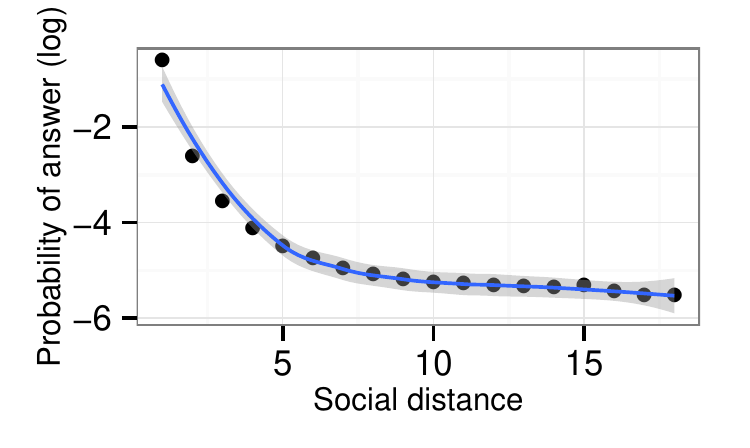}
\caption{Probability of getting answers from $h$-hop followers. Local polynomial regression fitting with $95$\% confidence interval area is also shown.}
\label{fig:prob_answers}
\end{figure}

Figure~\ref{fig:prob_answers} plots the geometric average of all these probabilities at a given distance as a function of social distance.
The figure confirms that the probability of receiving answers from $h$-hop followers decreases with social distance.

Therefore, the \emph{FF} network channels user attention, likely via its newsfeeds feature that sends updates to followers on the questions posted by the user. 
Does the same phenomenon hold true for abuse reports?

To answer this question we investigate both networks: along with the \emph{FF} which is an explicit network, we also investigate the activity network (\emph{AN}), which connects users based on their direct question-answer interactions.
For each (reporter, reportee) pair in the editorially-curated abuse reports, we calculated the shortest path distance between them in the social network and the activity network.
We compare our results with a null model that randomly assigns the abuse reports in our sample dataset to users in the two networks. 

\begin{figure}[htbp]
\centering
\includegraphics[height=4cm]{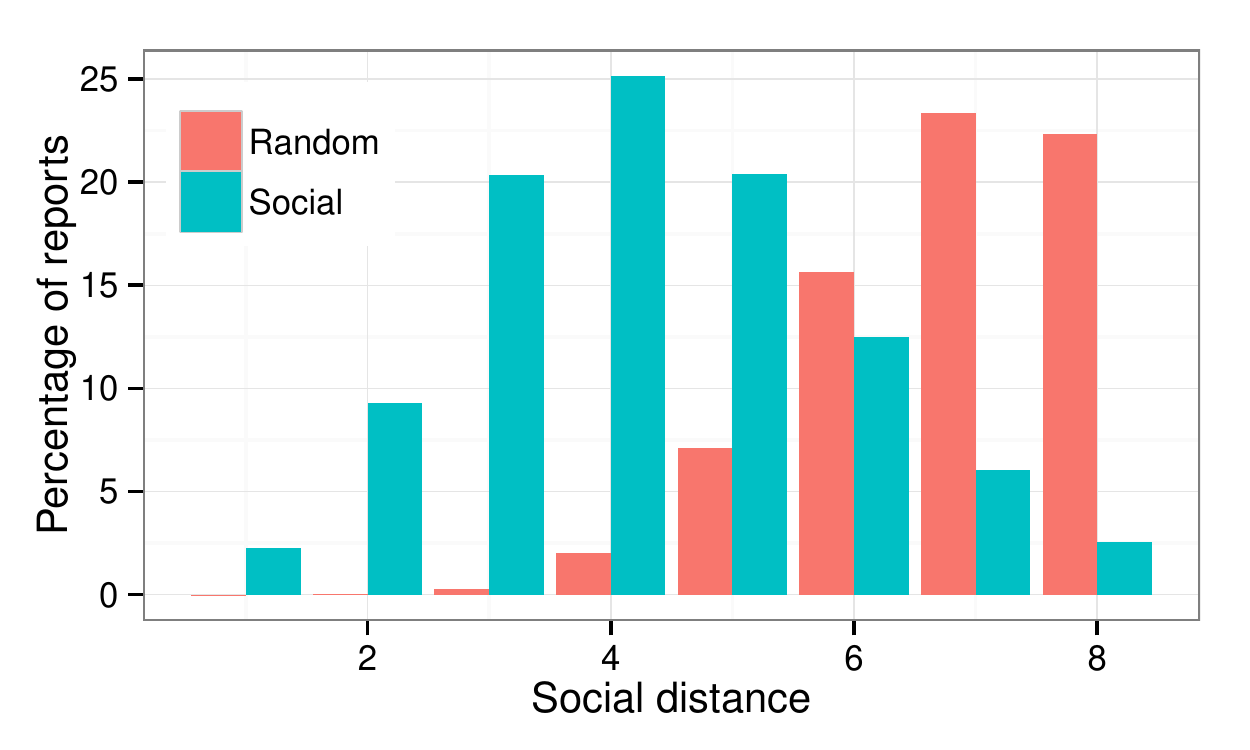}
\caption{Percentage of the abuse reports received by users from different distances in the social network, for the observed case and a random case.}
\label{fig:percent-social-flag}
\end{figure}

Figure~\ref{fig:percent-social-flag} shows the percentage of abuse reports users receive from close distances (up to 8 hops) for both (social and random) cases. 
About $75\%$ of the reports that users receive are from reporters located within $5$ social hops in the \emph{FF} network.
However, when reports are distributed randomly, about $9$\% are from within $5$ social hops and very few from within $3$ social hops.

\begin{figure}[htbp]
\centering
\includegraphics[height=3.9cm]{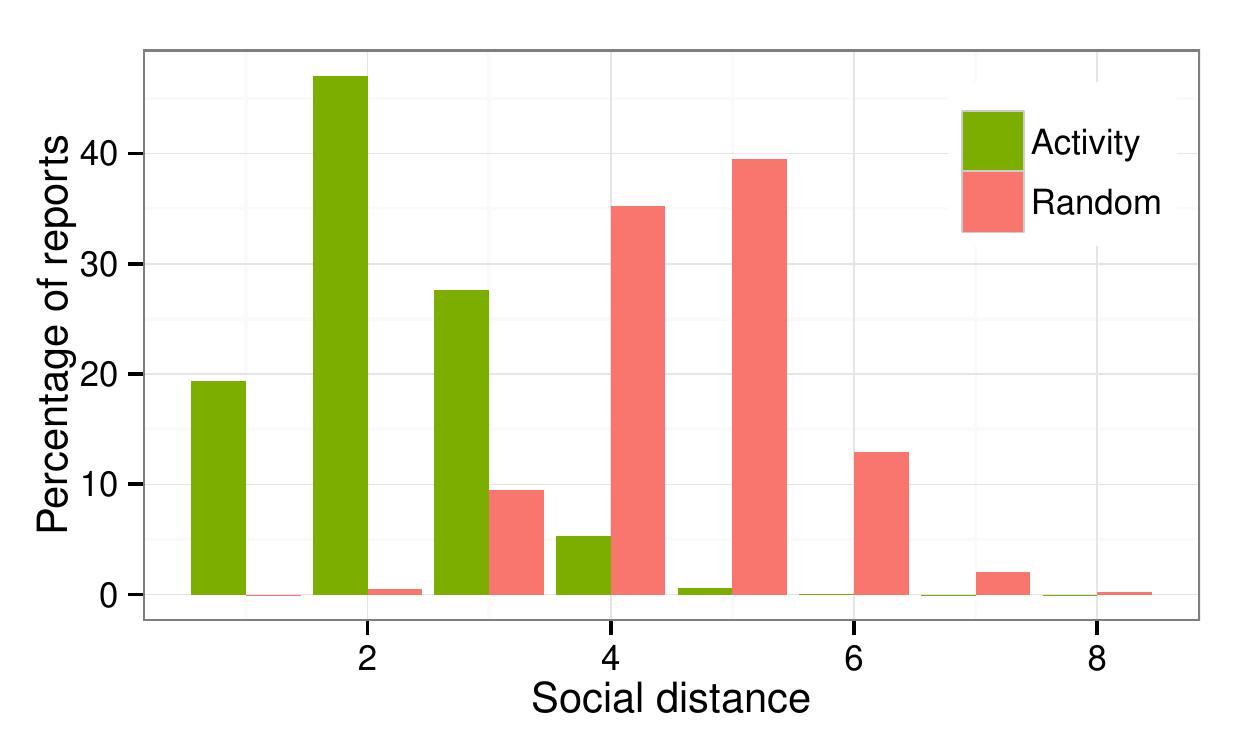}
\caption{Percentage of the abuse reports received by users from different distances in the activity network, for the observed case and a random case.}
\label{fig:percent-activity-flag}
\end{figure}

When comparing the percentage of abuse reports users receive with respect to distance in the \emph{AN} (Figure~\ref{fig:percent-activity-flag}), we notice that $94$\% of reports come from users within the first $3$ hops, which is significantly higher than the social network (about $32$\%).
We believe this is due to the high density of \emph{AN}: most of the nodes are reachable from others within a few hops.
However, even in this denser network, the null model has only about $10$\% of reports applied from within $3$ hops. 

To further quantify this phenomenon, we calculate the probability of being correctly flagged by users located at different network distances in the social and the activity network.
For a user $u$ and a social distance $h$, the probability of being flagged by followers at distance $h$ is the following:

\begin{equation}
p_{h}=  \frac{\textrm{\# of $u$'s followers at distance $h$ who flagged $u$}}{\textrm{\# of $u$'s followers at distance $h$}}
\end{equation}

\begin{figure}[htbp]
\centering
\includegraphics[height=2.90cm]{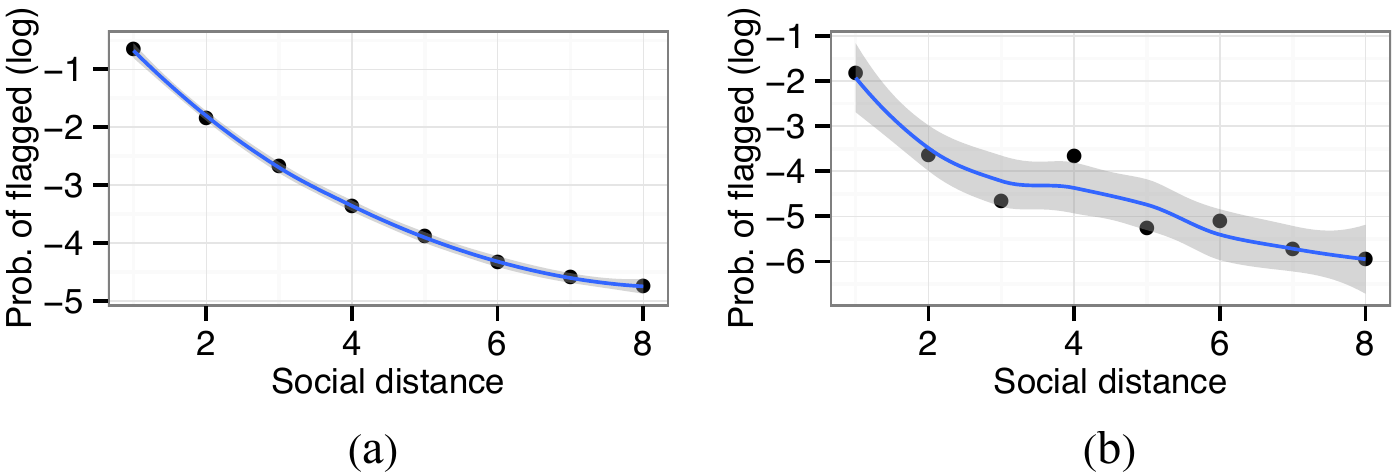}
\caption{Probability of being flagged by $h$-hop followers in the: (a) social network, and (b) activity network.  Local polynomial regression fitting with 95\% confidence interval area is also shown.}
\label{fig:prob_flag}
\end{figure}

Figure~\ref{fig:prob_flag} plots the geometric average of all probabilities at a given distance against the social distance for both networks.
As expected, the probability decreases with social distance in both the social and the activity networks.
The plot shows that users are likely to receive flags from others close to them in terms of social relationships and interactions.

These results confirm that the abuse reporting behavior is dominated by social relationships and interactions: users are reported for content abuse more from their close social or activity neighborhoods than from distant users.
The underlying reason is likely content exposure: a user's contents (questions/answers) are disseminated to nearby followers, thus they get higher exposure to that content compared to more distant users in the social graph.
Similarly, users who interact frequently with a user are more probable to view her contents and to report the inappropriate ones.

\subsection{Birds of a Feather Flock Together} \label{sec:assortativity}
Similarity fosters connection-- a principle commonly known as homophily, coined by sociologists in the 1950s.
Homophily is our inexorable tendency to link up with other individuals similar to us~\cite{mcpherson2001birds}.
In this section, we investigate whether homophily is also present in terms of deviance--that is, if deviant users tend to be close to each other in the social network.

One way to conclude about the homophily of a network is to compute  the attribute assortativity of the network~\cite{Newman2010Book}.
The assortativity coefficient is a measure of the likelihood for nodes with similar attributes to connect to each others.
The assortativity coefficient ranges between -1 and 1; a positive assortativity means that nodes tend to connect to nodes of similar attribute value, while a negative assortativity means that nodes are likely to connect to nodes with very different attribute value from their own.
If a network has positive assortativity coefficient, then it is often called assortative mixed by the attribute, otherwise called disassortative mixed.

In this work, we used question and answer-based deviant scores.
We considered each of the scores as an attribute and calculated the assortativity coefficient $r$ based on~\cite{newman2003mixing} for each type of deviance.   
The assortativity coefficients $r$ are shown in Table~\ref{tab:assortativitycoefficient} and are positive.
In~\cite{newman2003mixing}, Newman studied a wide variety of networks and concluded that social networks are often assortatively mixed (Table~\ref{tab:assortativitycoefficient} offers two such examples), but that technological and biological networks (e.g., World Wide Web $r=-0.067$,  software dependencies $r=-0.016$, protein interactions $r=-0.156$) tend to be disassortative. 
Comparing them quantitatively with the assortativity coefficients of the \emph{YA} network, we conclude that the \emph{YA} network is assortatively mixed in terms of deviance.
So, users having contacts with (low)high deviance scores will also have (low)high deviance scores.

\begin{table}
\caption{Assortativity coefficient $r$ for deviance scores in the \emph{YA} network.
Assortativity coefficients are also shown for other social networks from~\cite{newman2003mixing}.}
\centering
\scalebox{0.8}{
\begin{tabular}{c|c}
\hline
Yahoo! Answers			&	Other Social Networks\\ \hline
Question deviance $r=+0.11$	&	Mathematics coauthorship $r=+0.120$\\
Answer deviance $r=+0.13$	&	Biology coauthorship $r=+0.127$\\ \hline
\end{tabular}
}
\label{tab:assortativitycoefficient}
\end{table}

We next measure how similar the deviance scores of a user's contacts are with the user's, and how this similarity varies over longer social distances.
For this, we randomly sampled $100k$ users from the social network for each social distance ranking from 1 hop to 4 hops.
Let $U_h$ be the set of all the users ($100k$) selected for the social distance $h$.
We calculated the probability that user $u$'s $h$-hop contacts (with $u \in U_h$) will have the same deviance score as:
\begin{equation}
\resizebox{1.0\hsize}{!}{$
p_{u}=  \frac{\textrm{\# of $u$'s followers at distance $h$ with same deviance score}}{\textrm{\# of $u$'s followers at distance $h$}}
$}
\end{equation}

Rather than computing the exact similarity between a user and her follower's deviance scores, we focused on whether their difference is small enough to be dubbed as the same.
We considered two users' deviance scores are the same if their corresponding deviance score difference is less than a ``similarity delta''.
More specifically, $u$ will have \emph{about the same} deviance score with user $s$ located at distance $h$ if:

\begin{equation}
\left| deviance_u - deviance_s \right| < \delta
\end{equation}

The same technique was used for both types of deviance scores.We experimented with two values for $\delta$ equal to one or two standard deviations of the distribution of deviance scores in the network.
We report  the geometric average of all $p_u$ probabilities computed in each hop $h$.

\begin{figure}[htbp]
\centering
\includegraphics[scale=0.5]{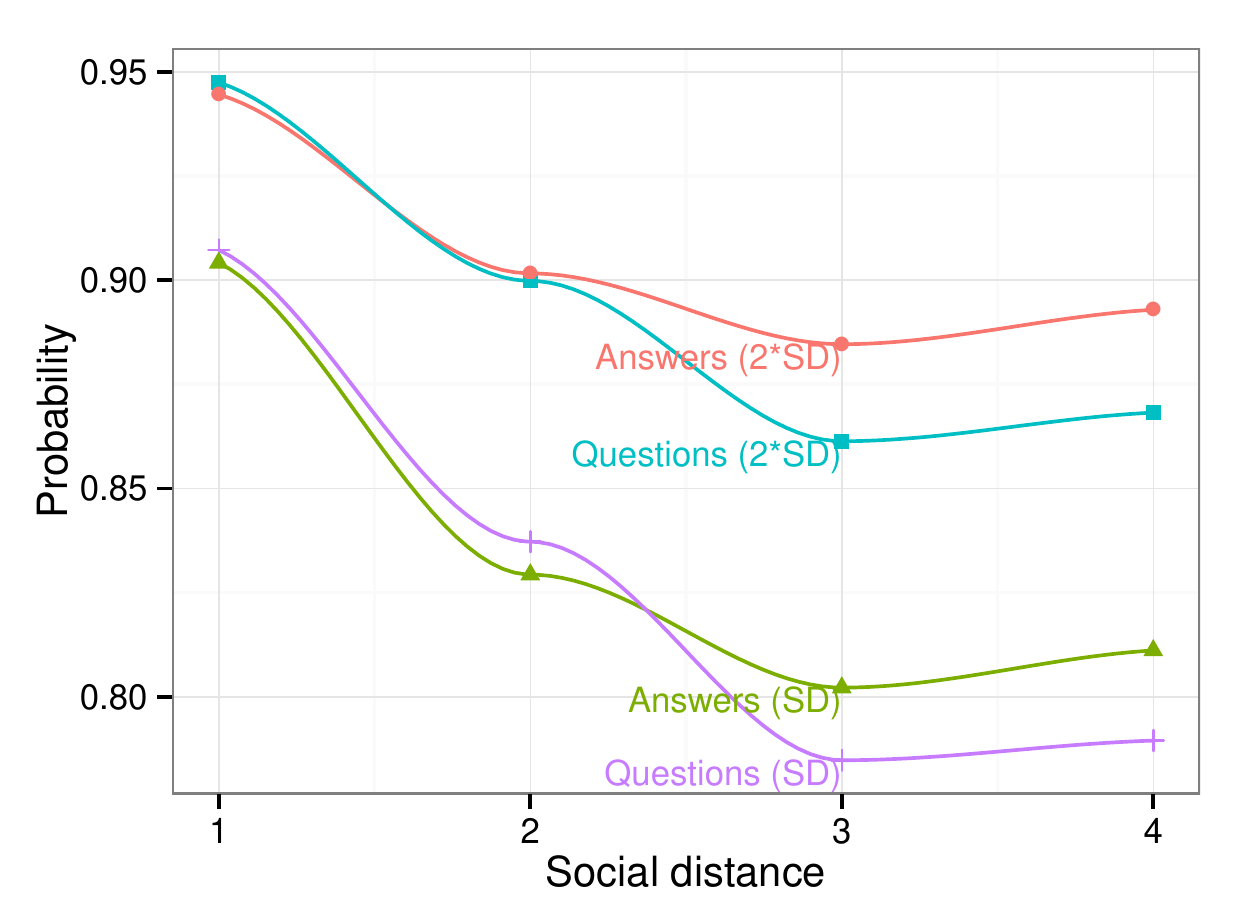}
\caption{Probability that a h-hop follower has the same deviant score to the user for $\delta = \sigma$ and $\delta=2\sigma$. SD: standard deviation.}
\label{fig:similarity_of_deviance}
\end{figure}

Figure~\ref{fig:similarity_of_deviance} shows the probability plots for both types of deviance, keeping similarity $\delta$ equal to one or two standard deviations.
Although different values of the $\delta$, the shapes of the figures are almost the same: up to 3-hops, the probability decreases gradually with the social distance.

\section{Suspended User Prediction}
\label{sec:classification}

Based on our previous analysis, we extract various types of features that we use to build predictive models.
We formulate the prediction task as a classification problem with two classes of users: fair and suspended.
Next, we describe the features used (Section~\ref{features}) and the classifiers tested (Section~\ref{setup}), and demonstrate that we are able to automatically detect fair from suspended users on \ya\ with an overall high accuracy (Section~\ref{classification-performance}).

\subsection{Features for Classification}\label{features}

Our predictive model has 29 features that are based on users' activities and engagements e.g., \textit{social}, \textit{activity}, \textit{accomplishment}, \textit{flag} and \textit{deviance}.
Table~\ref{tab:features} shows  the different categories of features used for the classification.
\textit{Social} features are based on the social network of the users, where \textit{Activity} features are based on community contributions in the form of questions and answers.
\textit{Accomplishment} features acknowledge the quality of user contribution (e.g., points, best answers).
\textit{Flag} summarizes the flags of a user (both received and reported). 
\textit{Deviance Score} features are the scores that we have computed based on users' flags and activities. 
Finally, \textit{Deviance Homophily} represents the homophilous behavior with respect to deviance.
Although most of the features are self-explanatory, below we clarify the ones which may not be.

\textbf{Reciprocity.} Reciprocity measures the tendency of a pair of nodes to form mutual connections between each other~\cite{garlaschelli2004patterns}.
Reciprocity is defined as follows:
\begin{equation*}
r= \frac{L}{L^*}
\end{equation*}
where $L$ is number of edges pointing in both directions and $L^*$ is the total number of edges.
$r=1$ holds for a network in which all links are bidirectional (purely bidirectional network), while a purely unidirectional network has $r=0$.

\textbf{Status.} Defined as the ratio of the number of a user's followers to her followees.

\textbf{Thumbs.} The difference between the number of up-votes and the number of down-votes a user receives for all her answers.

\textbf{Award Ratings.} The sum of the ratings a user receives for her best answers.

\textbf{Altruistic scores.} The difference between a user's contribution and his takeaway from the community.
For  altruistic scores, we consider \ya's point system, which awards two points for an answer, 10 points for a best answer, and penalizes five points for a question:
\begin{equation}
\begin{split}
 \textrm{Altruistic scores}_{u} &= f(contribution)-f(takeaway)\\
			&=2.0*A_u +10.0*BA_u-5.0*Q_u
\end{split}
\end{equation}
where $Q_u$ is the number of questions posted by $u$, $A_u$ is the number of answers posted by $u$, and $BA_u$ is the number of best answers posted by $u$.

\begin{table}[ht]
\caption{Different categories of features used for fair vs. suspended user prediction. We create a reciprocated network from the reciprocated edges. CC: clustering coefficient.}
\centering
\scalebox{0.8}
{
\begin{tabular}{l|c|l}
\hline
\textbf{Category} &\textbf{Number} & \textbf{Features}\\
   \hline
 \multirow{6}{*}{Social}& \multirow{6}{*}{6}&Indegree\\
    &&Outdegree\\
    &&Status\\
     &&Reciprocity\\
     &&Reciprocated networks degree\\
     &&Reciprocated networks CC\\
     \hline
 \multirow{4}{*}{Activity}& \multirow{4}{*}{4}&\#Questions\\
    &&\#Answers\\
    &&\#Flagged Questions\\
     &&\#Flagged Answers\\
          \hline
  \multirow{5}{*}{Accomplishment}& \multirow{5}{*}{5}&Points\\
    &&\#Best Answers\\
    &&Award Ratings\\
     &&Thumbs\\    
     &&Altruistic scores\\   
     \hline
         \multirow{8}{*}{Flag}& \multirow{8}{*}{8}&\#Question Flag Received\\
    &&\#Question Flag Received Valid\\
    &&\#Question Flag Reported\\
     &&\#Question Flag Reported Valid\\
     &&\#Answer Flag Received\\    
     &&\#Answer Flag Received Valid\\
    &&\#Answer Flag Reported\\
     &&\#Answer Flag Reported Valid\\    
     \hline
          \multirow{2}{*}{Deviance Score}& \multirow{2}{*}{2}&Question deviance score\\
    &&Answer deviance score\\
    \hline
          \multirow{4}{*}{Deviance Homophily}& \multirow{4}{*}{4}&Followers'  question deviance score\\
    &&Followers' answer deviance score\\
    &&Followees' question deviance score\\
     &&Followees' answer deviance score\\
    \hline
\end{tabular}
}
\label{tab:features}
\end{table}

\subsection{Experimental Setup and Classification}\label{setup}

In our dataset, the percentage of fair users (about $91$\%) are high compared to the suspended users (about $9$\%).
This leads to an unbalanced dataset.
Various approaches have been proposed in the machine learning literature to deal with the problem of unbalanced datasets.
We use the ROSE~\cite{menardi2014training} algorithm to create a balanced dataset from the unbalanced one.
ROSE creates balanced samples by random over-sampling minority examples, under-sampling majority examples or by combining over and under-sampling.
Our prediction dataset has $250k$ users with 60-40\% training--testing split. 
Using the under and over sampling technique of ROSE, we sample $150k$ users (fair and suspended each class has $75k$ users) to train  the classifier.
The testing set has $100k$ users, who are not present in the training dataset.
They are drawn randomly and fair vs. suspended ratio in the testing dataset is the same as the original \ya\ dataset.

We tested various classification algorithms, including Naive Bayes, K-Nearest Neighbors (KNN),  Boosted Logistic Regression, and Stochastic Gradient Boosted Trees (SGBT).
We use individual feature sets to investigate how successful each feature set is by itself, and then use all features for prediction.
For evaluation, we measure widely used metrics in classification problems: Accuracy, Precision, Recall and F1-score.
Table~\ref{settings} shows a summary of our experimental setup.

\begin{table}[ht]
\caption{Details of experimental setup.}
\centering
\scalebox{0.8}
{
\begin{tabular}{l|l}
\hline
\textbf{Dataset} 	&	$250k$ sampled users\\
\hline
\textbf{Class Balancing Alg.}		&	Random Over-Sampling Examples (ROSE)	\\
\hline
\textbf{Classifiers}				&	Stochastic Gradient Boosted Trees (SGBT)	\\
							&	Naive Bayes, Boosted Logistic Regression	\\
							&	K-Nearest Neighbors (KNN)				\\
							&	Support Vector Machines RDF				\\
\hline
\textbf{Feature Sets}				&	Social, Activity, Accomplishment			\\
							&	Flag, Deviance Homophily, All features		\\
\hline
\textbf{Train-Test Split}			&	$150k$ users training, $100k$ users testing	\\
\hline
\textbf{Cross Validation}			&	10-folds, repeated 10 times				\\
\hline
\textbf{Performance}				&	Accuracy, precision, recall, F1 score			\\
\hline
\end{tabular}
}
\label{settings}
\end{table}

\subsection{Classification Results and Evaluation}\label{classification-performance}

The performance results of various classifiers while using all features are shown in Table~\ref{classifiersPerformance2}.
The SGBT classifier outperforms other classifiers in all performance metrics.
This classifier offers a prediction model in the form of an ensemble of weak prediction models~\cite{friedman2002stochastic}.
In our setting, it achieves $82.61$\% accuracy in classifying fair vs. suspended users with a high precision ($96.94$) and recall ($83.52$).
The confusion matrix of the classifier is shown in Table~\ref{classifierPerformance1}.
The matrix shows that the SGBT classifier is able to correctly classify $83.52$\% of fair users and $73.39$\% of suspended users.

\begin{table}[ht]
\caption{Performance of  various classifiers using all available features.}
\centering
\scalebox{0.75}
{
\begin{tabular}{lcccc}
\hline
\textbf{Classifier Name} & \textbf{Accuracy} & \textbf{Precision} & \textbf{Recall}&\textbf{F1 Score}\\
\hline
Naive Bayes&47.21&96.93&43.34&59.89\\
Boosted Logistic Regression&71.61&96.62&71.28&82.03\\
KNN&73.81&96.41&73.97&83.71\\
SVM-RDF&75.92&95.62&77.06&85.34\\
SGBT&82.61&96.94&83.52&89.73\\
\hline
\end{tabular}
}
\label{classifiersPerformance2}
\end{table}

\begin{table}[ht]
\caption{Confusion matrix  for the SGBT classifier.}
\centering
\scalebox{0.80}
{
\begin{tabular}{ll|lll}
\hline
& & \textbf{Actual}\\
   \hline
 \multirow{2}{*}&\textbf{}&\textbf{Fair}&\textbf{Suspended}\\
    {\textbf{Predicted}}&\textbf{Fair}&\textbf{83.52}\%&26.60\%\\
    &\textbf{Suspended}&16.47\%&\textbf{73.39}\%\\
    \hline
\end{tabular}
}
\label{classifierPerformance1}
\end{table}

\begin{figure}[htbp]
\centering
\includegraphics[height=3.5cm]{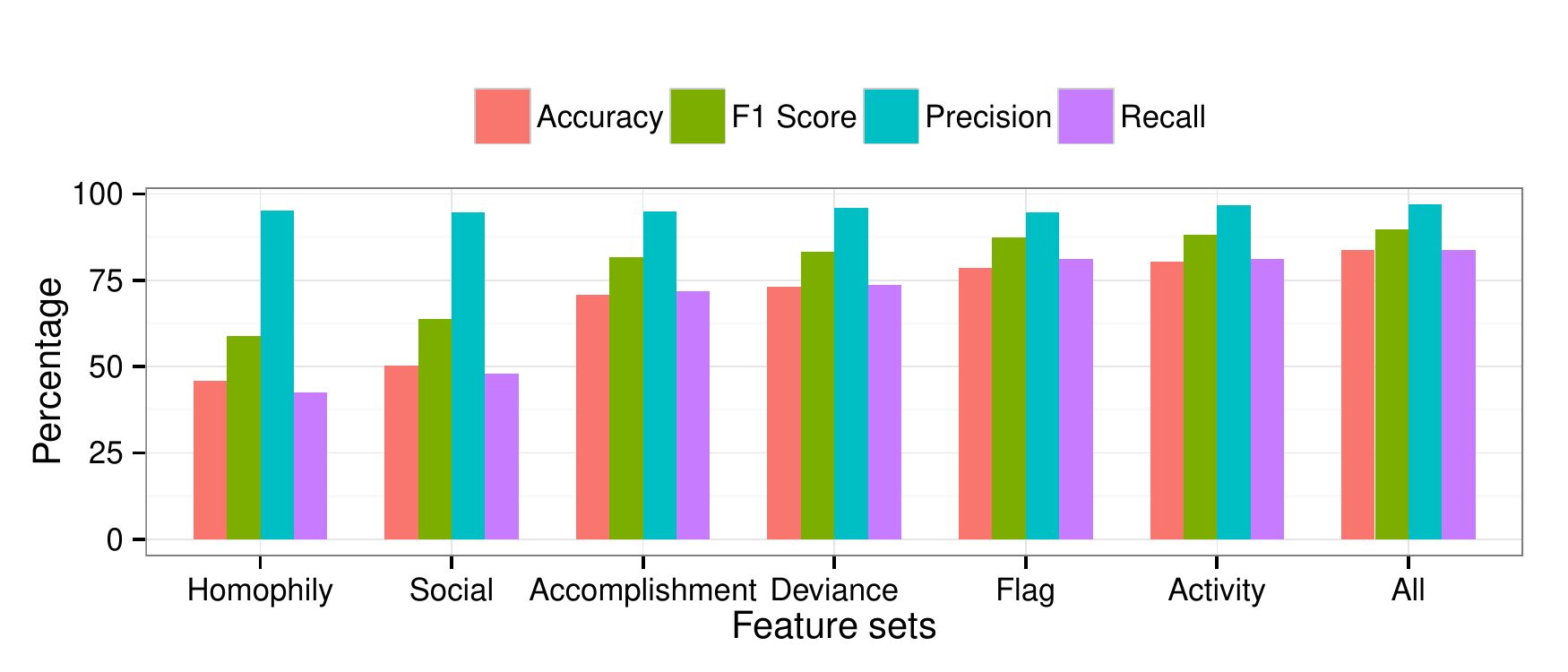}
\caption{Performance of the SGBT while classifying fair and suspended users using different feature sets.}
\label{fig:performanceFeatures}
\end{figure}

Figure~\ref{fig:performanceFeatures} shows the performance (accuracy, precision, recall and F1 score) of the models trained with different subsets of features using the SGBT classifier, which performs the best among the tested classifiers.
We observe that each feature set has a positive effect on the performance of the classifier across all performance metrics. 
This suggests that  all our feature sets are important for prediction.
Particularly, accomplishment, deviance, flags and activity features individually exhibit more than 70\% accuracy with good precision, recall and F1 score.
However, when all the features are used for classification, the performance metrics yield the best results, i.e., accuracy is improved by 4.11\% compared to activity features.

\begin{figure}[htbp]
\centering
\includegraphics[height=5cm]{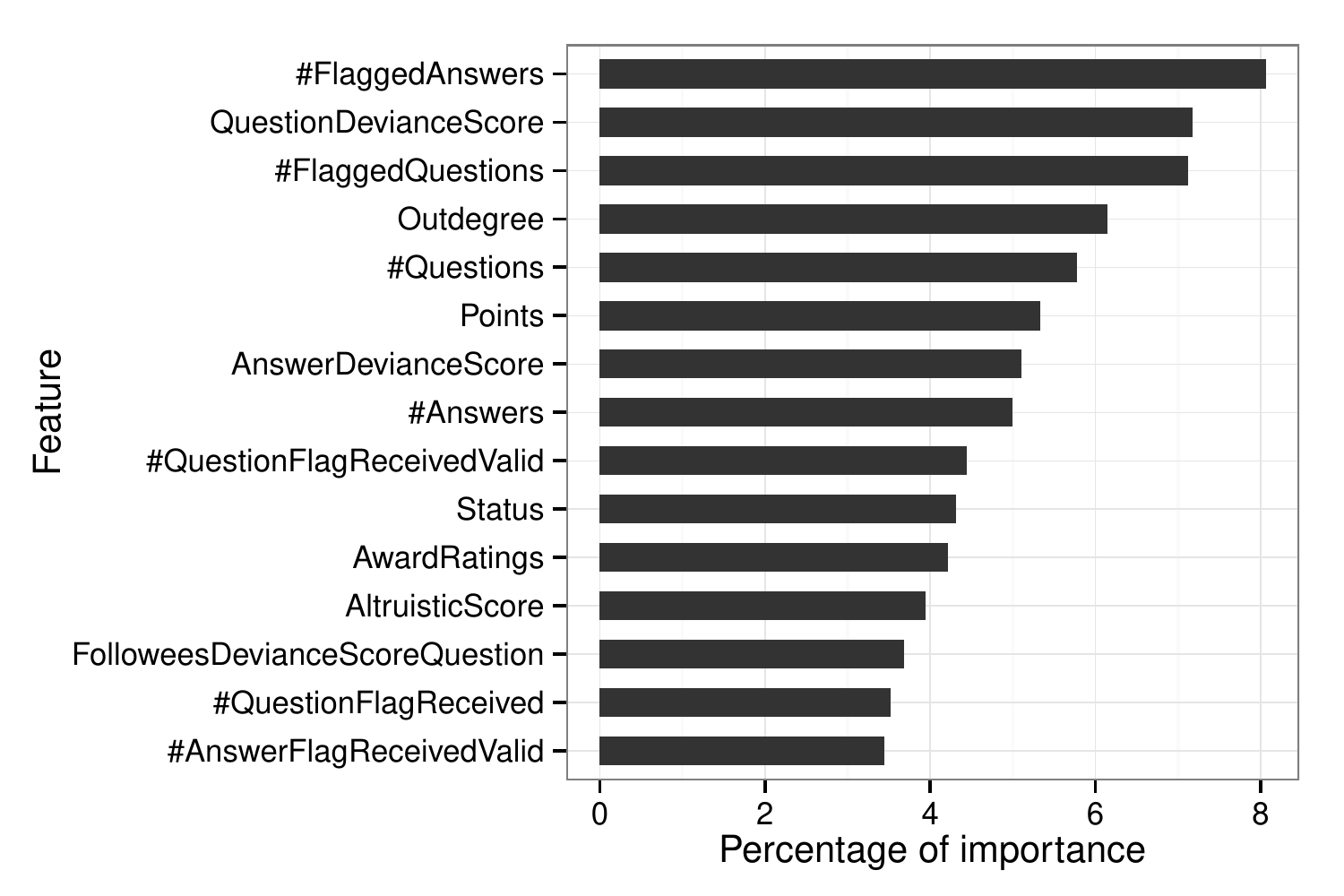}
\caption{Relative importance (out of 100, how much a feature is contributing) of top 15 features in classifying fair  and suspended users.}
\label{fig:feaureImportanceSuspendedvsFair}
\end{figure}

Figure~\ref{fig:feaureImportanceSuspendedvsFair} shows the most important features (top 15) in classification of fair vs. suspended users.
The model uses  a  backwards elimination feature selection method for feature importance. 
For each feature, the model tracks the changes in the generalized cross-validation error  and uses it as the variable importance measure.

We observe that the number of flagged content and deviance scores are the best predictors of fair and suspended users.
Also, at least one feature from all feature sets is within the top 15 features.
However, only activity and deviance score feature sets have all the features within the top 15 features.

\section{Summary and Discussion}
\label{sec:discussion}

This paper is an investigation of the flagging of inappropriate content in Yahoo Answers, a popular and mature Community-based Question-Answering platform.
Based on a sample of about 10 million flags in a population of about 1.5 million active users, our analysis revealed the following.

First, the use of flags is overwhelmingly correct, as shown by the large percentage of flags validated by human monitors.
This is an important learning for crowd sourcing, as it shows for the first time (to the best of our knowledge) that crowdsourced monitoring of content functions well in CQA platforms.
Moreover, although there are no explicit incentives (e.g., points) for flagging inappropriate content, users take the time to curate their environment.
In fact, 46\% of the users reported at least one abuse report, with the top abuse reporters flagging tens of thousands posts.

Second, we discovered that many users have collected a large number of flags, yet their presence is not deemed toxic to the community.
Even more, their contributions are engaging, which is certainly a benefit to the platform: the questions asked by the users who deviate from the norm (in terms of number of flags received for their postings) receive many more answers and from many more users than the questions posted by ordinary users or by users who later had their accounts suspended.
However, more content-based analysis is needed to understand how exactly the deviant users engage the community.
We posit that they might ask conversational, rather than informative, questions, as this behavior is shown to increase community engagement.

Third, we showed the importance of the follower-followee social network for channeling attention and producing answers to questions.
Less expected, perhaps, is the fact that this network also channels the attention of flaggers: we found that users in close social proximity are more likely to flag inappropriate content than distant users.
Social neighborhoods, thus, tend to maintain their environment clean.

Fourth, a significant problem in \ya\ is posed by the users who manage to avoid flagging, possibly by remaining at the outskirts of the social network.
This relative isolation in terms of followers and in terms of interactions probably allows such users to remain invisible.
They are likely caught by automatic spam-detection-like mechanisms and by paid human operators.
However, our empirical investigation showed that classifiers that use activity- and social network-based features can successfully identify fair and suspended ($40$\% of them are not flagged) users with accuracy as high as 83\%.

This work leads to various promising directions for future work.
Understanding what makes deviant users engaging can be helpful in designing strategies potentially applicable to a variety of communities.
Quantifying the equivalent behavior in terms of content abuse reporting and in terms of bad users on different online platforms can help understand the relative importance of different features for the success of the platform. 
And finally, characterizing (e.g., activity and social network centrality) the pro-social users who report abusive content may help identify such potential volunteers and appropriately incentivize them.

\section{Acknowledgments}
The work was funded by the National Science Foundation under the  grant CNS 0952420, and by the Yahoo's Faculty Research and Engagement Program.

\bibliographystyle{abbrv}
\bibliography{Bibtex}  %
\end{document}